\documentclass[12pt,preprint]{aastex}

\shorttitle{Polarmetric Images of Sgr A*} 
\shortauthors{Huang et al.}

\begin{document}


\title{Linearly and Circularly Polarized Emission in Sagittarius A*}


\author{Lei Huang\altaffilmark{1,3}, Siming Liu\altaffilmark{4}, Zhi-Qiang Shen\altaffilmark{1,2}, Mike J. Cai\altaffilmark{5}, Hui Li,\altaffilmark{4} and Christopher L. Fryer\altaffilmark{4} 
}

\altaffiltext{1}{Shanghai Astronomical Observatory, Chinese Academy of Sciences, Shanghai 200030, China; muduri@shao.ac.cn}
\altaffiltext{2}{Joint Institute for Galaxy and Cosmology (JOINGC) of ShAO and USTC, Shanghai 200030, China} 
\altaffiltext{3}{Graduate School of the Chinese Academy of Sciences, Beijing 100039, China} 
\altaffiltext{4}{Los Alamos National
Laboratory, Los Alamos, NM 87545}
\altaffiltext{5}{Academia Sinica, Institute of Astronomy and Astrophysics, Taipei 106, Taiwan}


\begin{abstract}

We perform general relativistic ray-tracing calculations of the transfer of polarized synchrotron radiation through the relativistic accretion flow in Sagittarius (Sgr) A*. Based on a two-temperature magneto-rotational-instability (MRI) induced accretion mode, the birefringence effects are treated self-consistently. By fitting the spectrum and polarization of Sgr A* from millimeter to near-infrared bands, we are able to not only constrain the basic parameters related to the MRI and the electron heating rate, but also limit the orientation of the accretion torus. These constraints lead to unique polarimetric images, which may be compared with future millimeter and sub-millimeter VLBI observations. In combination with general relativistic MHD simulations,  the model has the potential to test the MRI with observations of Sgr A*.

\end{abstract}

\keywords{
black hole physics --- Galaxy: center --- plasmas --- polarization ---
radiative transfer --- sub-millimeter}

\section{Introduction}

The compact radio source Sagittarius (Sgr) A* at the Galactic center is associated with a $\sim 4\times 10^6M_\odot$ supermassive black hole \citep{Sch02, Gh05} and is one of the best astrophysical sources for studying physics of black hole accretion. Since the discovery of active galactic nuclei and galactic X-ray binaries, it has been suggested that black hole accretion be responsible for the high luminosity and radiation efficiency of these sources. To release the gravitational energy of materials falling into black holes, the angular momentum must be transported to large radii in the accretion processes. The nature of the viscosity responsible for the angular momentum transport and energy dissipation is not well understood. \citet{SS73} proposed that the viscous stress is induced by turbulence and is proportional to the pressure of the accretion flow. 

MHD simulations have made the magneto-rotational-instability (MRI) the primary mechanism for generation of turbulence and viscous stress in accretion flows \citep{BH91, BH98}. Although there are indications that the viscous stress is proportional to the magnetic field energy density, the ratio of the magnetic field energy density to the gas pressure depends on the initial magnetic field configurations of these simulations \citep{Pes06}. \citet{KPL07} argue that, at least for some astrophysical accretion systems, the MRI may not be able to generate high enough viscosity to account for the observations. Nevertheless, the MRI is believed to play dominant roles in some black hole accretion systems. It has been studied extensively through MHD simulations and is awaiting observational test.

Sgr A* plays a crucial role in this aspect.  Since discovered in 1974 \citep{BB74}, the source has been observed extensively due to its association with the supermassive black hole \citep{MF01}. Radio imaging reveals a compact radio source with its intrinsic size marginally uncovered in the millimeter (mm)
band \citep{Bowe04, S05}. The centimeter (cm) spectrum can be fitted with a power-law function, and there is clear evidence for a mm and sub-mm spectral bump \citep{Falc98}. Radio ($ \le43$ GHz) emission of Sgr A* has unusual polarization characteristics with no detection ($<0.4\%$) of linear polarization (LP) and weak circular polarization (CP) of up to $\sim1.0\%$ in the quiescent-state \citep{Bowe02}. Weak LP ($\lesssim 1\%$) was detected recently at $43$ and $22$ GHz during radio outbursts \citep{Yus07a, Yus07b}. \citet{Macq06} reported a mean LP of $2.1 \pm 0.1\%$ at $3.5$ mm. In the sub-mm band, high LP of $\sim10\%$ has been routinely measured \citep{Aitk00, Marr07} without any firm indication of detectable CP \citep{Marr06}. 
These observations suggest that Sgr A* is powered by accretion of the supermassive black hole with the mm and sub-mm emission originating within 10 $R_S$, where $R_S\simeq 1.2\times 10^{12}[M/(4\times 10^6M_\odot)]$ cm is the Schwarzschild radius for a non-spinning black hole with mass $M$. The high LP implies a very low accretion rate \citep{QG00}. Based on the MRI, a small ($<10\ R_S$) accretion torus can account for the spectrum and polarization of the sub-mm emission \citep{Meli00, Meli01, Brom01}. \citet{Liu07}, hereafter L07, recently generalize the model to a two-temperature MRI driven accretion flow, which naturally explains the 90$^\circ$ flip between the electric vector position angles (EVPA) of the polarized sub-mm and near-infrared (NIR) emissions \citep{Ecka06, Meye07, Marr07}. Besides the basic parameters related to the MRI, the model introduces only one more parameter to characterized the not-well-understood electron heating processes. 

However, all models proposed so far have ignored the birefringence effects of synchrotron radiation propagating through the relativistic accretion flow in Sgr A* and have treated the general relativistic (GR) effects with simplified disk structures. The Faraday rotation of the birefringence effects can depolarize low frequency emission, and the GR light bending effect can complicate the transfer of polarized emission near the black hole.
Based on the model proposed by L07,  in this Letter we  present a more complete and self-consistent treatment of radiation from the relativistic accretion flow.The ratio of the magnetic field to gas pressure, emission spectrum, and polarization are very sensitive to the onset frequency of LP from the cm to mm bands, electron heating rate, and disk inclination angle, respectively.
A fitting to these observations leads to unique images of fractional LP and CP.
The combined effects of GR and birefringence produce distinct LP and CP features between the sub-mm and NIR bands. These predictions can be tested with future polarimetric VLBI observations. More importantly, in combination with MHD simulations, observations of Sgr A* can be used to test the MRI.
In \S\ \ref{method}, we briefly discuss our treatment of the transfer of polarized synchrotron radiation near a Schwarzschild black hole. Our main results are presented in \S\ \ref{result}. We draw conclusions and propose future work in \S\ \ref{cd}.

\section{Transfer of Synchrotron Radiation}
\label{method}

Following \citet{LL04}, hereafter LL04, we defined four vectors $\vec{p}, \vec{\epsilon}, \vec{\eta},$ and $\vec{\rho}$ to treat the transfer of polarized emission in a magnetized plasma. Here, $\vec{p}$ represents three normalized Stokes parameters $(Q, U, V)/I$, and $\vec{\epsilon}, \vec{\eta},$ and $\vec{\rho}$ correspond to three normalized emission coefficients $(\epsilon_Q, \epsilon_U, \epsilon_V) = (\varepsilon_Q, \varepsilon_U, \varepsilon_V) /I$, absorption coefficients $(\eta_Q, \eta_U, \eta_V)$, and Faraday conversion and rotation $(\rho_Q, \rho_U, \rho_V)$, respectively.
For the total specific intensity $I$, the corresponding emission and average absorption coefficients are indicated by $\varepsilon_I=\epsilon_I I$ and $\eta_I$, respectively. Then we have
\begin{eqnarray}
{{\rm d} I\over {\rm d} s} &=& - (\eta_I + \vec{\eta} \cdot \vec{p}-\epsilon_I)I\,,\\
{{\rm d} \vec{p}\over {\rm d} s} &=& -\vec{\eta} + (\vec{\eta} \cdot{} \vec{p})\vec{p}+\vec{\rho}\times\vec{p}+\vec{\epsilon}-\epsilon_I\vec{p}\,,
\end{eqnarray} 
where $s$ is the distance along a light ray. 

Synchrotron radiation is elliptically polarized with a dominant extraordinary emission component, whose electric field is perpendicular to both the magnetic field vector and the wave propagation direction, and the emission coefficients satisfy $\epsilon_I>(\epsilon_Q^2+\epsilon_U^2)^{1/2} \gg \epsilon_V$ \citep{B66, Melr71}. In the coordinate system with the ordinary and extraordinary components as the axes, the $U$ (or $Q$) components of $\vec{\epsilon}, \vec{\eta}$, and $\vec{\rho}$ vanish, and $\rho_Q$ (or $\rho_U$) and $\rho_V$ correspond to the Faraday conversion and rotation coefficients, respectively \citep*[][hereafter M97;KM98]{Melr97, KM98}. We use formulas of \citet{Melr71} to obtain $\epsilon_I$ and $\vec{\epsilon}$. Then $\eta_I$ and $\vec{\eta}$ can be derived from the Kirchhoff's law. The polarization of the natural wave modes of a magnetized plasma determines $\vec{\rho}$. Because there are no birefringence effects for the natural wave modes, $\vec{\rho}$ needs to be parallel to $\vec{\epsilon}$ (M97; KM98; LL04).  For the thermal plasmas we are interested in here, M97 has given expressions for the Faraday rotation coefficients in cold and relativistic limits. We extrapolate these results $\rho_V = (e^3 n B \cos{\theta_B}/\pi \gamma_c m_e^2 c^2\nu^2)[\gamma_c^{-1}(1-\ln{\gamma_c}/(2\gamma_c)) + \ln{\gamma_c}/(2\gamma_c)]$, where $e$, $m_e$, $\gamma_c m_e c^2/k_B$, $n$, $k_B$, $c$, $B$,  $\theta_B$, and $\nu$ are the electron charge, mass, temperature, density, Boltzmann constant, speed of light, magnetic field,  angle between the magnetic field and wave propagation direction, and  wave frequency, respectively. The Faraday conversion coefficient is then given by $\rho_V(\epsilon_U^2+\epsilon_Q^2)^{1/2}/\epsilon_V$.

We simulate the pseudo-Newtonian disk structure in a Schwarzschild metric with all quantities treated as measured by a locally static observer. Because the emission frequencies in the mm to NIR bands are much higher than the cyclotron and plasma frequencies of the thermal electrons in the accretion flow, the plasma effects on photon propagation discussed by \citet{BB03} can be neglected and all photons travel along the null geodesics.  To perform the above radiation transfer of polarized synchrotron emission with the ray-tracing code discussed in \citet{Huang07}, one only needs to take into account the rotation of EVPA along a null geodesic due to the space curvature \citep{Brom01}. The effects of magnetic field and GR have been taken care of in the radiation transfer equations \citep{BB04}.

MHD simulations of the MRI show that the magnetic field is dominated by its azimuthal component due to the shearing of the Keplerian accretion flow. Previous modeling of the sub-mm to NIR polarizations of Sgr A* assumed that the magnetic field does not have other components, see in \citet{Meli00, Brom01};L07. Here we consider two more realistic configurations: one with the magnetic and velocity field parallel to each other in the upper half of the disk and anti-parallel in the lower half, and the other with the magnetic field direction just reversed. 
The averaged Stokes parameters of these two configurations are used to fit observations. Sgr A* is highly variable with variation timescales decreasing with observing frequency. In mm and sub-mm bands, both 
flux and LP vary on hourly timescales \citep{Marr06}, while the accretion timescale near the black hole is a few days. Shorter timescale variations are likely associated with turbulent fluctuations in the accretion flow, which may be addressed with MHD simulations. The results of our stationary accretion model should be compared with measurements averaged over days. 

\section{Spectral Fitting and Polarimetric Images}
\label{result}

\begin{figure}[h]
\vspace{-0mm}
\begin{center}
\includegraphics[width=14.0cm]{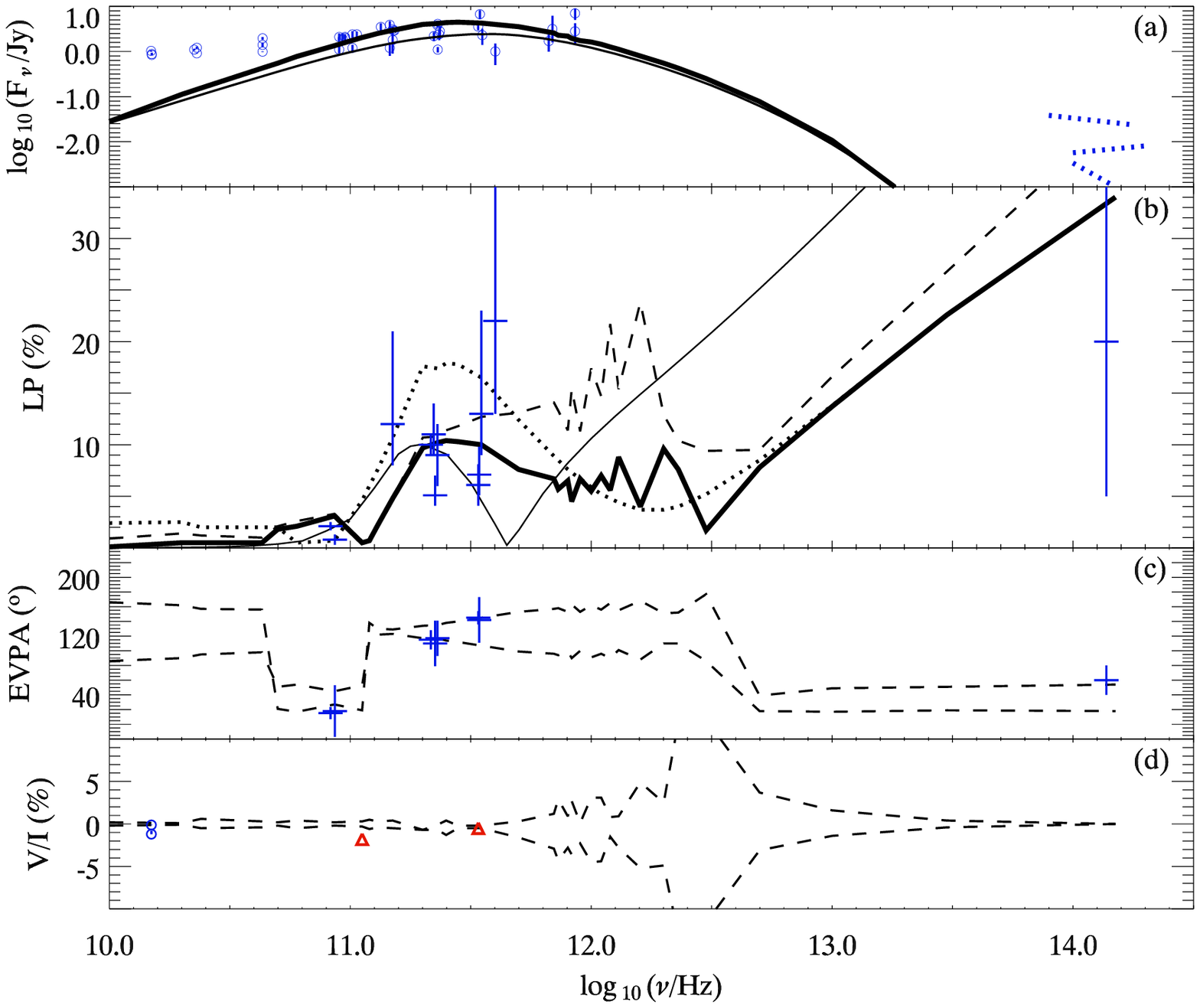}
\vspace{-5mm}\caption{Plots of the spectrum, LP, EVPA and CP of Sgr A* in panels (a),(b),(c), and (d), respectively. All the data points are from \citet{Liu07}, except the two CP measurements at 112 and 340 GHz (open triangles in panel (d)), which come from \citet{Bowe02} and \citet{Marr06}, respectively.  Although the NIR data are for flares, the optically thin emission is always dominated by emission from the blue-shifted side and the LP therefore can be compared with the model.
The thick solid lines in panels (a) and (b) correspond to spectrum and LP of the fiducial model and the thin solid lines represent the corresponding pseudo-Newtonian results.  The GR effects increase the flux density by a factor of $\sim2$ near the spectral peak. The dashed lines in panels (b),(c), and (d) show LP, EVPA, and CP of two magnetic field configurations, respectively. For LP, the two configurations give almost identical results so that only one is drawn. The dotted line in panel (b) shows the result with no birefringence effect, i.e. $\vec{\rho}=0$ . We predict strong CP variations near $1$ THz and three $90^\circ$ flips in the EVPA of the LP (see text for details).
\label{spec}}
\end{center}
\end{figure}

\begin{figure}[h]
\vspace{-0mm}
\begin{center}
\includegraphics[width=16.0cm]{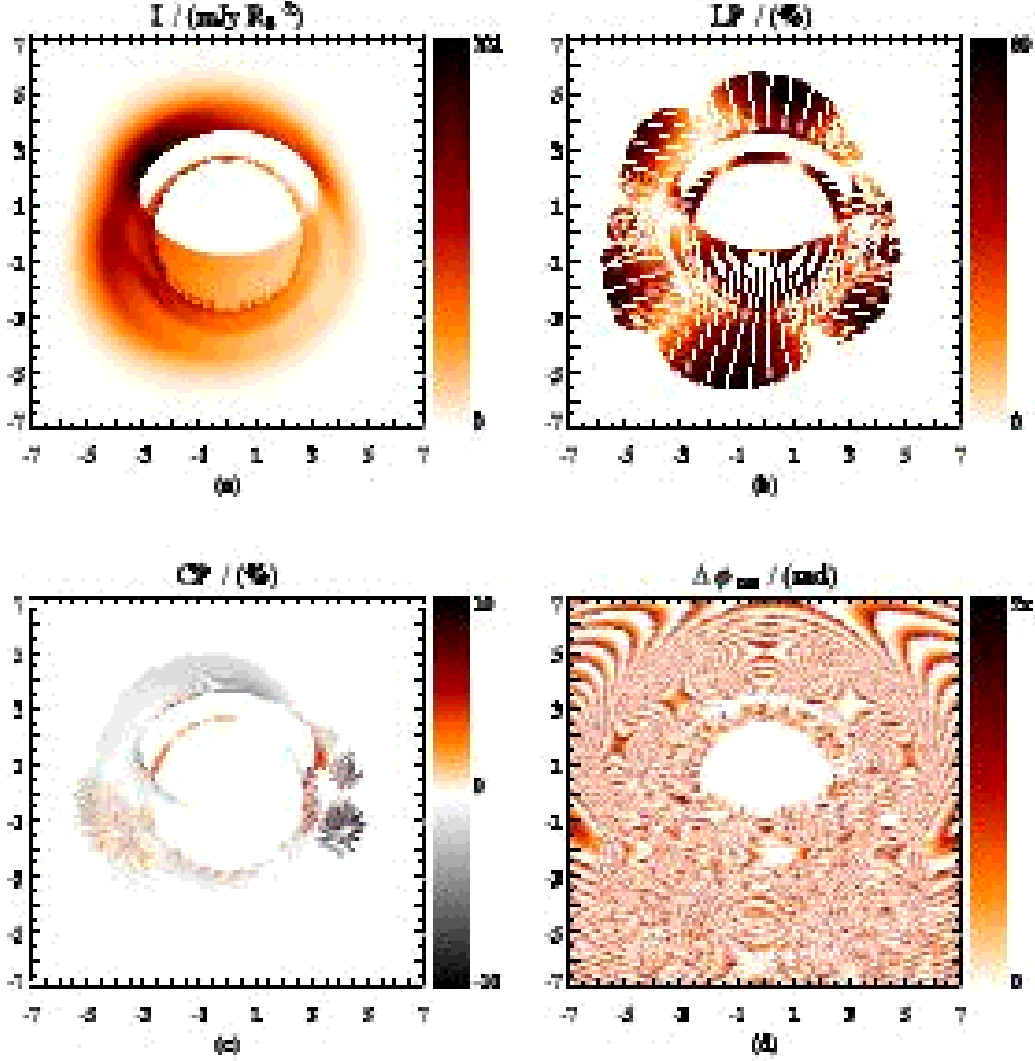}
\vspace{-5mm}\caption{Polarimetric images of the accretion flow in Sgr A* at 230 GHz. The length unit is  $R_S$ and the disk axis is in the vertical direction. Panel (a) shows the total intensity map. The left hand side is brightened by the Doppler boosting \citep{Brom01, BL06}. Panel (b) gives the map of the fractional LP with the EVPA and magnitude indicated by the white dashes and the color wedge, respectively. Panel (c) shows the fractional CP with the positive sign for right hand polarization (IEEE convention, see text for details). Panel (d) gives the residual of the integration of the Faraday conversion coefficient along the rays divided by $2\pi$, with which the strips in the LP and CP images are clearly correlated.
\label{image}}
\end{center}
\end{figure}

The thick solid lines in Fig.\ref{spec}(a) and (b) represent the best fit to the spectrum and polarization data of Sgr A*. The data points are summarized in L07, 
except the two CP measurements at $112$ GHz \citep{Bowe02} and $340$ GHz \citep{Marr06}.  We fit the data around mm to sub-mm bump, including the EVPAs at $86$ GHz averaged over five epochs in 2004 \citep{Macq06}, and at $230$ GHz that remained unchanged during observations from 2002 October to 2004 January \citep{Bowe05}.  In our fitting, neither the NIR measurement \citep{Ecka06} nor the detected LP at $22$ and $43$ GHz \citep{Yus07a,Yus07b} is used simply because these are related to flaring activity.  However, the EVPA at $2.2\mu$m remained unchanged from 2004 to 2006 and may be explained by a temporary disk with toroidal magnetic field \citep{Ecka06, Meye07}. The corresponding five model parameters are the ratio of  viscous stress to magnetic field energy density $\beta_\nu =0.7$, the ratio of magnetic field energy density to gas pressure $\beta_p =0.4$, the electron heating rate indicated by the dimensionless parameter $C_1 =0.44$, the accretion rate $\dot{M} =0.95 \times 10^{-8} M_\odot$ yr$^{-1}$, and the inclination angle of the accretion flow $i=40^\circ$.
The most prominent GR effect is the amplification of the emission area due to the light bending near the black hole (see Fig.\ref{image}). At higher frequencies the disk becomes optically thin and lower frequency emission is produced at relatively larger radii, so the flux density enhancement by the GR effects is significant near the spectral peak and vanishes in the low and high frequency limits as can be seen by comparing with the thin solid line of the pseudo-Newtonian spectrum.  

The GR effects also change the LP. The thin solid line in Fig.\ref{spec}(b) shows the corresponding pseudo-Newtonian result. The LP is significantly suppressed ($\lesssim 1\%$) by birefringence effects at low frequencies.
Another radio component introduced by L07,  which is presumably associated with an outflow, may fit the cm CP and flux density spectra \citep{BF02, Zhao04} and explain the polarization observed during radio outbursts \citep{Yus07a, Yus07b}. 
The dotted line shows the GR calculation without the birefringence effects, i.e. $\vec{\rho}=0$. We see that there is a weak level of LP ($\sim2-3\%$) at low frequencies. The birefringence effects reduce the LP in mm and sub-mm bands by up to $50\%$. 
At very high frequencies, the birefringence effects are weak and the results approach that of the fiducial model. As shown by L07,  the flux density is very sensitive to the electron heating rate $C_1$, which determines the electron temperature profile. However, the fitting to the emission spectrum doesn't give a unique set of model parameters. With the additional LP measurements, the best fit model parameters can be fixed. This is because the synchrotron emissivity is proportional to $B^2 n$ while the Faraday rotation measure is proportional to $B n$, for a $B^2 n$ constrained by the emission spectrum, the Faraday rotation increases with the decrease of $B$, which is determined by $\beta_p$. Therefore the onset frequency of LP increases with the decrease of $\beta_p$. 
The two magnetic field configurations give almost identical LP but different EVPA and CP. So we plot only one LP [dashed line in Fig.\ref{spec}(b)] but two sets of EVPA and CP [dashed lines in Fig.\ref{spec}(c) and (d)].  The average of Stokes parameters over the two configurations reduces the LP. One may use the difference between the average and individual configurations to estimate the variability of the corresponding quantities.

The most interesting prediction of our model is oscillations of LP and CP with frequency $\sim 1$ THz (see Fig.\ref{spec} (b) and (d)). This is due to the combined effects of GR and birefringence.  Since $\vec{\rho}\propto \nu^{-2}$, the birefringence effects are negligible $\gtrsim 1$ THz. The modulation of  LP and CP by the birefringence is amplified by the GR effect $\sim 1$ THz, where $\int |\vec{\rho}| {\rm d} s$ is on the order of $2 \pi$.  The LP oscillates with the amplitude and period decreasing due to rapid increase of the Faraday rotation angle as frequency decreases, in contrast to the pseudo-Newtonian and no-birefringence GR cases.
This effect also gives rise to high CP of $>5\%$ for each individual magnetic field configurations.  Future polarization observations should be able to easily test these predictions.  
The model predicted intrinsic EVPA flips by $\sim 90^\circ$ three times from cm to NIR bands. The flip at $\sim 50$ GHz is related to the fact that the red-shifted side of the Keplerian flow 
becomes transparent and strongly polarized in the direction perpendicular to thedisk axis.  At $\sim 100$ GHz,  the near and far sides also become transparent and dominate the LP. The EVPA flips to be roughly aligned with the disk axis. At $\sim 3$ THz,  the polarized emission is dominated by the blue-shifted side and the EVPA flips again. Without introducing an external Faraday rotation measure, 
the range limited by the two configurations fits the EVPA measurements well.
The corresponding position angle of the disk axis is $\sim 120^\circ \pm 30^\circ$, which is different from that inferred from the NIR flare observations \citep{Meye07} and the pseudo-Newtonian model (L07).  The fitting to the EVPA can be improved with an external rotation measure \citep{Macq06, Marr07}, which also brings the EVPA of the disk axis in agreement with the pseudo-Newtonian model. 
Similar to the pseudo-Newtonian case, the LP and EVPA flips are very sensitive to the inclination angle of the disk. 
In practice, cases with inclination angles ranging from $30^\circ$ to $50^\circ$ can fit the spectrum and LP within error bars. The best fit EVPA of $40^\circ$ is $10^\circ$ smaller than the pseudo-Newtonian result.  Nearly edge-on $(i\ge60^\circ)$ cases overestimate the LP in the sub-mm band and fail to produce a $90^\circ$ EVPA flip from sub-mm to NIR bands because the near side of the torus dominates the emission. 

Fig.\ref{image} shows the intrinsic polarimetric images of the fiducial model at 230 GHz with the disk axis in the vertical direction. Panels (a), (b), (c), and (d) are the images of the total specific intensity $I$, fractional LP ($\sqrt{Q^2 + U^2} / I$), fractional CP ($V / I$), and the residual ($\Delta \phi_{\rm{res}}$) of the Faraday rotation angle $\int(\rho_Q^2 + \rho_U^2)^{1/2}{\rm d} s$ divided by $2\pi$, respectively.  Results shown in Fig.\ref{spec} are obtained by integrating the Stokes parameters over the disk to compare with spatially unresolved observations.
Doppler boosting enhances $I$ on the left hand side of Fig.\ref{image}(a). The white dashes in Fig.\ref{image}(b) represent the intrinsic EVPA of each ray.   
In Fig.\ref{image}(c) positive (red) and negative (black) values of CP correspond to right and left hand polarization, respectively\footnote{The convention used for CP ($V/I$) is consistent with the IEEE definition which has been adopted by the IAU and used by many observations, i.e., right hand polarization corresponds to the electric vector of an approaching wave rotating counter-clockwise.}.
The strips in Fig.\ref{image}(b) and (c) are clearly correlated with the Faraday conversion shown in Fig.\ref{image}(d). 
In our case, the Faraday conversion coefficient is always much greater than the Faraday rotation coefficient, and its integrations along the rays are much larger than $2\pi$ at 230 GHz, the Stokes $V$ therefore changes its sign over one period of EVPA rotation (KM98; LL04). 
The strips will be less distinct in a realistic turbulent flow. The observed images will always be smoothed by finite spatial resolutions and interstellar scattering effects \citep{S05}.  However, rough contours may be resolved by future polarization-sensitive VLBI observations since the scattering effect is less important in sub-mm band.  
The black hole shadow may be detected in the $I$ image, as evidence for the GR theory. We also expect high levels of LP with the EVPA mostly in the radial direction, as evidence for toroidal component dominated magnetic fields.

\section{Discussion and Conclusion}
\label{cd}

Tremendous efforts have been put to image shadows caused by the strong gravity of black holes \citep{KG06, S05, Bowe04}. Sgr A* has been the best candidate for such studies. The polarization measurements of Sgr A* have put strong constraints on the physical processes near the black hole. 
Based on the two-temperature MRI driven Keplerian accretion model, which has only five parameters, we calculate the synchrotron emission from the accretion flow with both the GR and birefringence effects taken into account self-consistently. The model can reproduce the spectrum and polarization of Sgr A* in mm and shorter wavelength bands and further predicts distinct LP and CP at 
$\sim 1$ THz. Polarimetric images of the relativistic accretion flow of the best fit model are made, which can be verified with future polarization-sensitive VLBI observations. 

The MRI has become the leading mechanism for the generation of viscous stress required to drive the accretion flow of plasmas into black holes. The theory has been well established through both theoretical analyses and numerical investigations. The model presented here makes it possible for the first time to put the theory into observational tests. If the scaling laws for the MRI discovered by \citet{Pes06} are applicable to the accretion flow in Sgr A*, the fiducial model parameter $\beta_p=0.4$ requires the Alfv\'{e}n velocity induced by the large scale vertical magnetic field be $\sim20$ times smaller than the sound velocity. With this large scale magnetic field configuration, one can carry out GR MHD simulations for Sgr A* \citep{NG06}. The outflows from these simulations should not only reproduce low frequency radio flux, CP, and source size measurements, but also induce a Faraday rotation measure external to the disk studied here, which can improve the fit to the observed EVPAs in mm and sub-mm bands. NIR and X-ray flares and the high variability of this source, which cannot be studied with the current model, may also be addressed with such simulations \citep{B01, Gil06, H07, Meye07}.

\acknowledgments

This work was supported in part by the NNSFC (grants 10573029, 10625314, 10633010, and 10733010), and
the Knowledge Innovation Program of the CAS (KJCX2-YW-T03), and sponsored by 
the Program of Shanghai Subject Chief Scientist (06XD14024), and 
the National Key Basic Research Development Program of China (2007CB815405), and under the auspices of the US DOE by its contract W-7405-ENG-36 to the LANL. ZQS acknowledges the support by the One-Hundred-Talent Program of the CAS.


\clearpage

\end{document}